# From Deposition Stress to Surface Reactivity: Strain-Dependent Hydrogen Evolution on Sputtered Platinum Thin Films

*Sabrina Baha[1], Alejandro E. Perez Mendoza[2], Leonardo H. Morais[1], Aleksander Kostka[3], Shivam Shukla[2], Ellen Suhr[1], Andre Oliveira[2†], Annika Gatzki[1], Henrik H. Kristoffersen[4], Jan Rossmeisl[4], Corina Andronescu[2]*, Alfred Ludwig[1,3]**

[1] Material Discovery and Interfaces, Ruhr-University Bochum, Bochum, Germany

[2] Chemical Technology III, University Duisburg-Essen, Duisburg, Germany

[3] Center for Interface-Dominated High Performance Materials, (ZGH) Ruhr-University Bochum, Bochum, Germany

[4] Center for High Entropy Alloy Catalysis, University of Copenhagen, Copenhagen, Denmark

ABSTRACT

Strain has emerged as a promising approach for tuning electrocatalytic properties, yet its role in sputter-deposited thin films remains poorly understood. In this work, magnetron-sputtered platinum (Pt) thin films with different stress states were prepared by varying the sputter pressure.

The resulting changes in microstructure, residual strain, and hydrogen evolution reaction (HER) activity were investigated using complementary characterization techniques and density functional theory (DFT) calculations. Structural analysis reveals a transition of (111)-textured Pt thin films from dense and smooth films at low pressures, to more porous microstructures with increased roughness at higher pressures. Electrochemical measurements show that films deposited at low sputter pressure exhibit the highest HER activity, while higher sputter pressures lead to reduced activity despite increased surface area. DFT calculations demonstrate that lattice strain alters hydrogen adsorption energetics and surface coverage on Pt(111), providing a mechanistic explanation for the observed activity trends. Overall, the results highlight that HER activity in sputtered Pt thin films is governed by the interplay of residual strain, microstructure, and hydrogen coverage.

## INTRODUCTION

Platinum (Pt) is a benchmark electrocatalyst for energy conversion systems as it is one of the most efficient catalysts for the hydrogen evolution reaction (HER) [1,2] as well as oxygen reduction reaction (ORR) in acidic electrolyte [3,4]. Its high intrinsic activity originates from its electronic structure, which provides near-optimal adsorption energies for key reaction intermediates [5]. Since the electrocatalytic performance of a material is strongly determined by the electronic configuration of its surface atoms, strategies that tune the electronic structure, such as alloying, increasing chemical complexity in high-entropy materials or modifying the local atomic coordination at the surface, play a central role in catalyst design [6,7]. Among these approaches, strain engineering has emerged as a powerful concept, as lattice distortions can directly shift the d-band center of Pt and thereby modulate adsorption energetics [8-11].

Strain in electrocatalytic materials may arise from lattice mismatch between the substrate and the material of interest, structural defects, or externally applied mechanical deformation. Numerous studies have demonstrated that both tensile and compressive strain can significantly influence HER and ORR kinetics by altering the electronic states of metal catalysts [12-18]. For example, He et al. demonstrated dynamic control of surface strain in ultrathin Pt shells, achieving lattice strain ranging from approximately −5.1% (compressive) to +5.9% (tensile). The best-performing catalyst, exhibiting ~2.8% tensile strain, reached a specific activity of 22.71 mA/cm$^2$ at −0.07 V vs. RHE, corresponding to a 12.3-fold enhancement relative to Pt/C [19]. Similarly, tensile strain arising from geometric constraints in Pd icosahedra resulted in nearly two-fold higher HER current density (−4002 vs. −1929 mA/cm$^2$ at −0.87 V vs. RHE), accompanied by a corresponding doubling of turnover frequency (9178 vs. 4427 s$^{-1}$), further underscoring the quantitative coupling between lattice distortion and HER performance [9].

Density functional theory (DFT) plays an important role in understanding electrocatalytic activity by using adsorption energies as descriptors for reaction trends [20]. The hydrogen adsorption free energy ($\Delta G_H$) is widely used as a key activity descriptor for HER, where optimal electrocatalytic performance is expected to be obtained for adsorption free energies close to 0 eV [20,21]. On Pt(111), hydrogen adsorption energies have been shown to exhibit a pronounced dependence on surface coverage due to lateral adsorbate–adsorbate interactions. These interactions are reflected in coverage-dependent adsorption energies. Previous studies have shown that hydrogen adsorption energies on Pt(111) vary by up to 0.2 eV as the coverage increases to one monolayer, reflecting the progressive weakening of the Pt-H interaction due to repulsive interactions at higher surface coverages [22, 23]. Strain provides another systematic means to shift $\Delta G_H$ on Pt(111). Both tensile and compressive strain modify $\Delta G_H$ through strain-induced changes in the surface electronic structure [24,25]. Martínez-Alonso et al. reported that tensile strain strengthens hydrogen adsorption on Pt(111), whereas compressive strain weakens it. Across strain ranges from 3% compressive strain to 8% tensile strain, adsorption energies shift up to 0.25 eV toward more negative values.

While strain-engineered model systems clearly demonstrate the pronounced sensitivity of electrocatalytic activity to subtle changes in atomic spacing, comparatively little attention has been paid to the residual stress states that naturally emerge during the fabrication of thin-film electrocatalysts on substrates. In sputter-deposited films, intrinsic stress is inherently linked to the typical non-equilibrium growth conditions. Parameters such as sputter pressure and the kinetic energy of impinging species govern adatom mobility, grain coalescence, defect incorporation, and film densification [26–29], ultimately determining whether compressive or tensile residual stress develops during film growth. Therefore, films prepared under different sputter conditions can exhibit substantially different residual stress states, which manifest as corresponding residual strain and ultimately impact electrocatalytic performance. Although previous studies have shown that sputter parameters influence morphology, preferred orientation, and occasionally electrochemical behavior of Pt-based thin films [30], a systematic and quantitative correlation between deposition-induced residual strain and electrocatalytic activity has not yet been established. Establishing this relationship is critical, as the resulting residual strain constitutes an inherent yet often uncontrolled parameter that may significantly influence electrocatalytic performance, not only of Pt but also possible future compositionally complex catalysts.

In this work, we investigate the relationship between residual strain in sputter-deposited Pt thin films and their HER activity in acidic electrolyte. Pt thin films were deposited using different Argon (Ar) sputter pressures, enabling systematic tuning of the residual stress state and the resulting residual strain. The magnitude of the residual stress was quantified using a microcantilever-based curvature measurement technique (digital holographic microscopy, DHM) [31], providing direct access to the biaxial in-plane stress. The films were further characterized by atomic force microscopy (AFM), X-ray diffraction (XRD), and transmission electron microscopy (TEM) to resolve surface morphology, crystal structure, and microstructural features across multiple length scales. HER activity was evaluated in acidic electrolyte using voltammetric measurements performed using a scanning droplet cell (SDC), with current densities normalized to the electrochemically active surface area (ECSA). To provide theoretical atomistic insight into the experimentally observed trends, DFT calculations were performed to evaluate the influence of biaxial strain on hydrogen adsorption energetics on Pt(111). By quantitatively correlating deposition-process-induced residual strain with DFT-derived hydrogen adsorption energetics and experimentally measured HER activity, we establish a direct multiscale link between thin-film growth conditions, elastic lattice distortion, surface electronic structure, and electrocatalytic performance. We demonstrate that variations in sputter pressure induce changes in biaxial residual strain of ~0.7%, accompanied by variations in overpotential of up to ~77% across the investigated pressure range. These findings highlight sputter deposition as a viable strategy to tune residual strain in Pt thin films and thereby modulate their electrocatalytic response.

EXPERIMENTAL METHODS

## Fabrication of Pt thin films

Pt thin films were deposited using a magnetron sputter system (CMS, DCA instruments) equipped with a radio frequency (RF) power supply. The base pressure before introducing Ar was below 6.7 × $10^{-6}$ Pa. A Pt target (99.99% purity) was used for film deposition, while a tantalum (Ta) target (99.95% purity) was used as an adhesion layer. All depositions were carried out under identical geometric conditions with a target to substrate distance of 10 mm. The RF power was fixed at 60 W for all Pt films to ensure comparability between samples. No intentional substrate heating was applied during deposition. The Pt films were deposited onto different substrates, including silicon (Si) wafer pieces with a 500 nm thermal oxide layer ($Si/SiO_2$), $Si/SiO_2$ substrates with a photolithographically cross-patterned structure for thickness measurements, and commercial Si microcantilever stress sensor chips (Micromotive GmbH) containing 20 $\mu$m thick cantilevers with a native oxide surface layer for residual stress analysis. During deposition, substrates were rotated at 10 rpm to ensure thickness homogeneity. Prior to Pt depositions, all substrates were coated with a Tantalum (Ta) adhesion layer of approximately 10 nm thickness. The Ta target was first pre-cleaned at 150 W for 300 s, followed by Ta deposition at 150 W for 175 s at a sputter pressure of 0.67 Pa under Ar (99.9999%) atmosphere. Pt films with a nominal thickness of 60 nm were then deposited, yielding a total film thickness of approximately 70 nm, including the Ta adhesion layer. The Pt target was pre-cleaned at 50 W for 300 s prior to each deposition. The sputter pressure was systematically varied to tune the residual stress state. Pressure was controlled via a throttle valve, with increased Ar flow at higher sputter pressures to ensure stable operating conditions. The film thickness was adjusted via deposition time based on calibrated sputter rates. Thickness determination is described in the following section. The corresponding sputter parameters are summarized in Table 1.

**Table 1.** Deposition parameters for Pt thin films deposited by RF magnetron sputtering at constant RF power (60 W). Film thickness values represent the total thickness of the Ta adhesion layer and Pt layer.

| Sample | Sputter Pressure (mTorr) | Sputter Pressure (Pa) | Ar Flow ($\times10^{-6}$ $m^3\ s^{-1}$) | Sputter Rate ($nm\ s^{-1}$) | Total Thickness (Ta + Pt) (nm) |
|---|---|---|---|---|---|
| p1 | 1 | 0.13 | 3.33 | 0.057 | 69 ± 3 |
| p2 | 2 | 0.27 | 3.33 | 0.052 | 69 ± 3 |
| p4 | 4 | 0.53 | 3.33 | 0.052 | 68 ± 2 |
| p5 | 5 | 0.67 | 3.33 | 0.053 | 69 ± 3 |
| p6 | 6 | 0.80 | 3.33 | 0.056 | 69 ± 4 |
| p7 | 7 | 0.93 | 3.33 | 0.056 | 70 ± 3 |
| p8 | 8 | 1.07 | 3.33 | 0.052 | 69 ± 1 |
| p10 | 10 | 1.33 | 3.33 | 0.062 | 72 ± 4 |
| p12 | 12 | 1.60 | 5.00 | 0.058 | 74 ± 3 |
| p20 | 20 | 2.67 | 6.67 | 0.067 | 67 ± 3 |
| p25 | 25 | 3.33 | 8.33 | 0.066 | 53 ± 1 |

**Film Thickness Determination**

The total thin film thickness (Ta + Pt) was determined with a mechanical profilometer (AMBIOS XP-1). To enable accurate thickness measurement, a well-defined step edge (cross-pattern) was created by a photolithographic lift-off process prior to deposition. Briefly, selected areas of the substrate were protected with a photoresist layer, while the remaining surface was exposed for metal deposition. After sputtering, the photoresist was removed using acetone, followed by rinsing with isopropanol, thereby lifting off the overlying metal and generating sharp step edges between coated and uncoated regions. The film thickness was then obtained by scanning across this step using profilometry. For each sample, at least five measurements were performed at different positions across the substrate to ensure statistical reliability. The values reported in Table 1 correspond to the mean thickness and standard deviation of these measurements. The measured total thickness was consistent with the nominal Pt thickness and the Ta adhesion layer described above. For sample p25, cross-sectional TEM indicated a lower local film thickness than initially obtained by profilometry. A subsequent profilometer recalibration and repeated measurements confirmed a systematic offset for this sample. The corrected thickness value was therefore used for all thickness-dependent analyses.

**Residual Stress Measurement**

The residual stress in the thin films was determined using Digital Holographic Microscopy (DHM) through measurement of the substrate curvature of the microcantilever stress sensor chip, enabling extraction of the film-averaged biaxial in-plane residual stress from cantilever deflection. Throughout this work, this film-averaged biaxial in-plane residual stress is referred to as residual stress unless stated otherwise. The average residual stress in the film, $\sigma_f$, was calculated from the measured radius of curvature $r$ using the Stoney equation:

$$\sigma_f = \frac{M_s h_s^2}{6 h_f r} \tag{1}$$

where $M_s = \frac{E_s}{1-\nu_s}$ is the biaxial modulus of the substrate with $E_s$ as the Young's modulus of and $\nu_s$ as the Poisson's ratio of the substrate, respectively, $h_s$ and $h_f$ are the substrate and film thicknesses, respectively [32]. Tensile residual stress corresponds to concave cantilever bending (positive curvature), whereas compressive residual stress corresponds to convex bending (negative curvature). For each sample, at least ten individual DHM measurements were performed on a single stress chip comprising six cantilevers, and the reported stresses represent mean values. The corresponding biaxial in-plane elastic strain, $\varepsilon_{\parallel}$, was obtained from: $\varepsilon_{\parallel} = \sigma_f / M_{111}$. Based on the pronounced (111) texture observed by XRD, the biaxial modulus $M_{111}$ of Pt(111) was determined using the anisotropic elastic expression for cubic crystals under equi-biaxial plane stress conditions as given by Freund and Suresh. Using the single-crystal elastic constants of Pt at 300 K ($C_{11}$ = 346.7 GPa, $C_{12}$ = 250.7 GPa, and $C_{44}$ = 76.5 GPa) yields $M_{111} \approx$ 337 GPa [33,34].

The total residual stress can be expressed as the sum of intrinsic and thermal contributions, $\sigma_{tot} = \sigma_i + \sigma_{th} + \sigma_{ext}$. The intrinsic stress component arises from non-equilibrium film growth processes during sputtering, including atomic peening, defect incorporation, grain coalescence, and microstructural evolution [35, 36]. The thermal stress component results from the mismatch in thermal expansion coefficients between film and substrate upon cooling or heating relative to the deposition temperature [37, 38]. Although deposition was performed nominally at room temperature, moderate substrate heating during magnetron sputtering cannot be excluded. Literature reports substrate temperature increases of several tens of degrees Celsius during magnetron sputtering in the absence of active cooling [35]. For the Pt/Si system, this would correspond to a thermal strain of approximately 0.009%, which is small compared to the experimentally determined residual strain values and can therefore be considered negligible. Since no external mechanical load was applied during curvature measurements and the thermal contribution is minor, the measured residual stress predominantly reflects the intrinsic growth stress of the films.

**Surface Roughness Analysis**

AFM was used to quantify the surface roughness of all Pt thin films. For each sample, an area of $1 \times 1\ \mu m^2$ was analyzed to assess potential correlation between morphological and electrocatalytic performance. Measurements were performed using a scanning probe microscope (FastScan, Bruker USA), operated in PeakForce Tapping mode with ScanAsyst control. The FastScan A S-probe with a nominal spring constant of 18 N $m^{-1}$ was used. The acquired height images were processed using NanoScope Analysis software version 1.8, Bruker USA. Prior to roughness evaluation, a first order plane correction was applied to remove sample tilt. No additional filtering procedures were applied. The arithmetic mean roughness $R_a$ and the root mean square roughness $R_q$ were determined from the full $1 \times 1\ \mu m^2$ topography image. The error was estimated as the standard deviation of five 300 x 300 $nm^2$ regions of interest within the same image.

**X-ray Diffraction**

XRD was performed to investigate the phase constitution and crystallographic structure of the samples using a D8 Discover diffractometer, Bruker USA, equipped with a VANTEC 500 two-dimensional detector and a Cu Kα microfocus radiation source (50 W, IμS, $\lambda = 0.15406$ nm). Five frames were acquired with an exposure time of 60 s per frame over a 2θ range of 30° - 80° in Bragg-Brentano geometry. The resulting two-dimensional diffraction patterns were converted into one-dimensional diffractograms using DIFFRAC.EVA software. The incident X-ray beam size was defined by a 1 mm collimator with a divergence below 0.007°. The peak layer accuracy of the instrument was ±0.02 in 2θ°. The interplanar spacing $d_{111}$ of the Pt(111) reflection was determined from the peak position using Bragg's law. The out-of-plane strain was calculated as $\varepsilon_{\perp} = (d_{111} - d_{111,0})/d_{111,0}$, where $d_{111,0} = 0.2265$ nm was used as the stress-free reference value [39].

**Transmission Electron Microscopy**

Thin cross-sectional samples (lamellae) for TEM analysis were prepared from selected thin films (p1, p2, p7, p8, p12, and p25), which were chosen to represent different residual strain and corresponding HER activities. Lamellae were prepared using a focused ion beam (FIB) system (FEI Helios G5 CX) operated at 30 kV. To protect the surface of the material, a carbon layer was deposited first, using an electron beam, and then a thicker layer, using an ion beam. The lift-out, transfer, and initial thinning of the sample down to 500 nm were performed at 30 kV. From then, the accelerating voltage of the ion beam was gradually reduced during the thinning process: first, thinning from 500 nm to 200 nm at 16 kV, and finally, thinning to electron transparency was performed at 8 kV. TEM characterization was conducted using a JEOL JEM-ARM200F (NEOARM) microscope operated at 200 kV. Prior to acquisition of selected area electron diffraction (SAED) patterns, the camera length of the instrument was calibrated using an evaporated aluminum film standard. All SAED patterns were acquired using identical condenser, objective and projector lens currents to minimize the influence of the experimental setup on the determination of the lattice parameters. The lattice parameters were determined from SAED patterns by manually selecting the diffraction center and determining peak positions based on their full width at half maximum. Interplanar spacings were calculated from the measured peak positions and converted to lattice parameters assuming an fcc structure. For each SAED pattern, multiple reflections ((111), (200), (220), (311), and (222)) were evaluated, and the resulting lattice parameters were averaged. To estimate the experimental uncertainty, four independent evaluations were performed on a SAED pattern of comparatively low quality, resulting in a standard deviation of 0.002 nm. This value is taken as a conservative estimate of the uncertainty of the lattice parameter determination.

**Electrochemical Measurements**

The HER activity of the Pt thin films in 0.1 M $HClO_4$ was evaluated based on linear sweep voltammograms (LSVs) recorded using an SDC (Sensolytics GmbH, Germany) equipped with a circular tip opening of 0.0019 $cm^2$. Proper sealing was ensured by monitoring the applied force, which was set to −400 mN after surface contact. The electrochemical setup consisted of a three-electrode configuration with the Pt thin film as the working electrode, an Ag|AgCl|3 M KCl as the reference electrode (Sensolytics GmbH), and a Pt wire as the counter electrode. A 0.1 M $HClO_4$ solution (prepared from $HClO_4$ 70%, from Sigma) was used as the electrolyte and continuously purged with Ar (99.999% purity, Airliquide). Purging was initiated 20 min prior to the experiment to remove dissolved $O_2$. Additionally, an Ar flow was directed around the capillary opening to minimize oxygen diffusion into the droplet during measurements. Three consecutive measurements were performed on each Pt surface, and the electrolyte was automatically exchanged before each new measurement area was evaluated. The SDC setup allowed automated measurements at multiple positions on each sample under comparable experimental conditions, and enables a systematic comparison of Pt thin films deposited under different sputter conditions.

Initially, the open circuit potential (OCP) was recorded for 150 s. Electrochemical impedance spectroscopy (EIS) was performed at the OCP using a 10 mV amplitude with 10 points per decade in the frequency range of 100 kHz to 1 kHz. The uncompensated resistance ($R_u$) was determined from the high-frequency intercept of the Nyquist plot with the real impedance axis. Cyclic voltammograms (CVs) were recorded from -0.26 to 0.63 V vs. Ag|AgCl|3 M KCl at a scan rate of 150 mV/s to determine the electrochemical active surface area (ECSA) of the Pt films. The charge associated with the hydrogen desorption was determined by integrating the anodic sweep between -0.22 and 0.08 V vs. Ag|AgCl|3 M KCl after double layer correction. The ECSA was calculated using a conversion factor of 210 μC/cm$^2$ [40]. Subsequently, LSVs were recorded from 0.14 V vs. Ag|AgCl|3 M KCl toward negative potential at a scan rate of 100 mV/s until a current of 50 μA was reached, corresponding to approximately 26 mA/cm$^2$. This stop criterion was applied to limit hydrogen bubble formation that could destabilize the droplet cell.

For data presentation, potentials were converted to the reversible hydrogen electrode (RHE) scale according to $\mathrm{E_{RHE}} = \mathrm{E_{Ag|AgCl|3M\ KCl}} + 0.210 + 0.059 \times \mathrm{pH} - \mathrm{i}R_u$, where $\mathrm{E_{Ag}}$|AgCl|3M KCl is the applied potential vs. Ag|AgCl|3 M KCl, 0.210 V is the standard potential of the reference electrode at 25 °C, *i* is the measured current and $R_u$ is the uncompensated resistance determined by EIS. For ECSA determination, the integrated hydrogen desorption charge after double layer capacitance correction was divided by 210 μC/cm$^2$.

**DFT Calculations**

DFT calculations were performed using the revised Perdew–Burke–Ernzerhof (RPBE) [41] exchange–correlation functional within the projector augmented wave (PAW) framework, as implemented in GPAW [42]. A plane wave cutoff energy of 400 eV and a 4 × 4 × 1 Monkhorst–Pack k point mesh were employed. All geometries were optimized until residual forces below 0.05 eV Å$^{-1}$. The Pt(111) surface was modelled as a 3 × 3 × 5 atomic supercell using the Atomic Simulation Environment (ASE) [43]. Periodic boundary conditions were applied, and a vacuum region of 2 nm was introduced perpendicular to the slab to avoid spurious slab–slab interactions. The two bottom layers were held fixed, while the remaining layers were relaxed until residual forces were below 0.05 eV/Å. Biaxial tensile and compressive strain were applied along the x and y directions by varying the lateral lattice parameters according to $a = a_0(1 + s)$, where $a_0$is the equilibrium lattice constant of Pt(111) and $s$ denotes the applied strain. Tensile (s > 0) and compressive (s < 0) strain were imposed in increments of 0.5%, up to a maximum magnitude of 1%. Hydrogen adsorption energetics were evaluated within the computational hydrogen electrode framework (CHE), enabling first principles evaluation of adsorption thermodynamics. The adsorption energy was calculated as

$$\Delta E_H = E_{Pt-H} - E_{Pt} - \frac{1}{2}E_{H_2} \tag{2}$$

where $E_{Pt-H}$ is the total energy of the slab with adsorbed hydrogen, $E_{Pt}$ is the energy of the clean surface, and $E_{H_2}$ is the energy of molecular hydrogen in the gas phase. Free energy of adsorption, $\Delta G_H$, was obtained according to

$$\Delta G_H = \Delta E_H + \Delta E_{ZPE} - T\Delta S_H \tag{3}$$

where $\Delta E_{ZPE}$ corresponds to the difference in zero-point energy between adsorbed hydrogen and hydrogen in the gas phase, and $T\Delta S_H$ represents the entropic contribution. Gas phase entropy values were taken from Atkins [44], while adsorbed species contributions followed Nørskov et al. [20]. This results in a total free energy contribution of 0.24 eV per hydrogen atom at room temperature. The CHE formalism was used to incorporate the effect of an applied potential $U$ vs. RHE, such that the adsorption free energy becomes $\Delta G_H(U) = \Delta G_H - eU$. To estimate the HER activity, $\Delta G_H$ was used as an activity descriptor in accordance with the Sabatier principle. The HER activity was evaluated using

$$j_k = Ae^{\frac{-(|\Delta G_H - \Delta G_{opt}| + eU)}{k_B T}} \tag{4}$$

where $\Delta G_{opt} = 0$ corresponds to the ideal adsorption free energy, $e$ is the elementary charge, is the Boltzmann constant, $T = 298$ K is room temperature, and $A = -0.01$ a.u. is a scaling constant chosen to reproduce experimentally observed trends [45-47]. Potentials were evaluated in the range from –0.8 to 0 V versus RHE.

RESULTS AND DISCUSSION

**Dependence of thin film morphology and strain on sputter pressure**

The sputter-deposited Pt thin films exhibit a clear transition from compressive to tensile stress with increasing sputter pressure, consistent with previous observations [48,49]. The corresponding residual strain exhibits a volcano-shaped dependence on sputter pressure (Figure 1). At the lowest sputter pressure of 0.27 Pa, the thin film shows a compressive strain of –0.034% (corresponding to a stress of –429 MPa). Increasing the sputter pressure to 1.07 Pa leads to a maximum tensile strain of 0.119% (481 MPa), resulting in a total residual strain range of ~0.15%. The change of the residual strain with sputter pressure is consistent with the well-established mechanism of energetic densification at low pressures (high incident particle energy) and stress relaxation at higher pressures due to reduced intercolumnar coupling caused by grain boundary porosity (low incident particle energy) [50]. At low sputter pressures, the mean free path of sputtered species is large, allowing Pt atoms and reflected neutral Ar to reach the substrate with substantial kinetic energy. This promotes atomic peening, where energetic bombardment leads to local densification [51,52]. The dominance of atomic peening under these conditions is consistent with the formation of a dense Zone T morphology [53]. At Ar pressures of around 1 Pa, the Pt thin film exhibits a pronounced tensile-stress maximum. This transition originates from the reduced kinetic energy and directionality of the sputtered Pt species as the working-gas pressure increases. At higher sputter pressures, increased gas-phase scattering is expected to randomize the incidence angles of the arriving Pt atoms, thereby reducing the lateral flux coherence and promoting a growth mode in which the film's grains increasingly develop into separated columnar structures. This resulting columnar morphology contains separate grain boundaries that interrupt the atomic network and reduce the film's ability to maintain coherency, thereby enabling relaxation of deposition-induced stress. Simultaneously, the reduced kinetic energy of the incoming species likely limits surface diffusion, causing adatoms to remain close to their initial impact position. Under these conditions, surface roughness increases and shadowing effects become increasingly important, leading to preferential growth of elevated regions. Such pressure-dependent changes in microstructure and residual stress are widely observed in sputtered metals [30, 54-57].

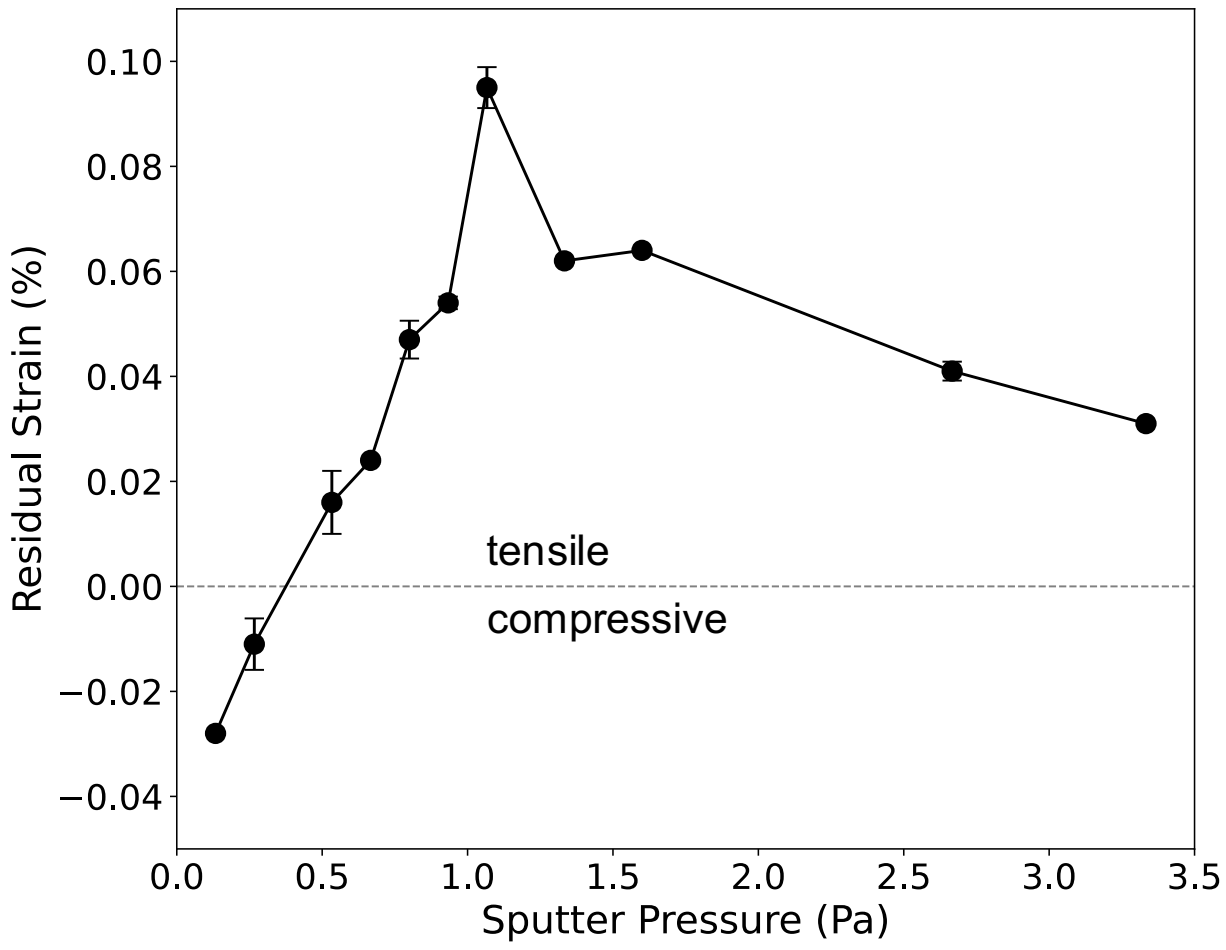

**Figure 1.** Relationship between sputter pressure and residual strain in Pt thin films. Strain values were calculated from residual stress measurements obtained by DHM. The horizontal line at y = 0 represents the unstrained state, and error bars indicate the uncertainty associated with the DHM-derived strain values.

This microstructural transition is accompanied by systematic changes in surface morphology, as evidenced by the AFM measurements shown in Figure 2. The lowest surface roughness is observed for the thin film deposited at 0.27 Pa (p2; $R_{a,0.27}$ = 0.63 ± 0.01 nm, $R_{q,0.27}$ = 0.79 nm ± 0.01 nm), whereas the sample grown at 3.33 Pa exhibits the highest surface roughness (p25; $R_{a,3.33}$ = 1.80 nm ± 0.07 nm, $R_{q,3.33}$ = 2.28 nm ± 0.08 nm). At an Ar pressure of 1.6 Pa, a pronounced increase in surface roughness is observed, which coincides with the formation of a columnar microstructure that is no longer laterally connected at the surface. AFM images of the remaining samples are provided in the Supporting Information (Figure S1).

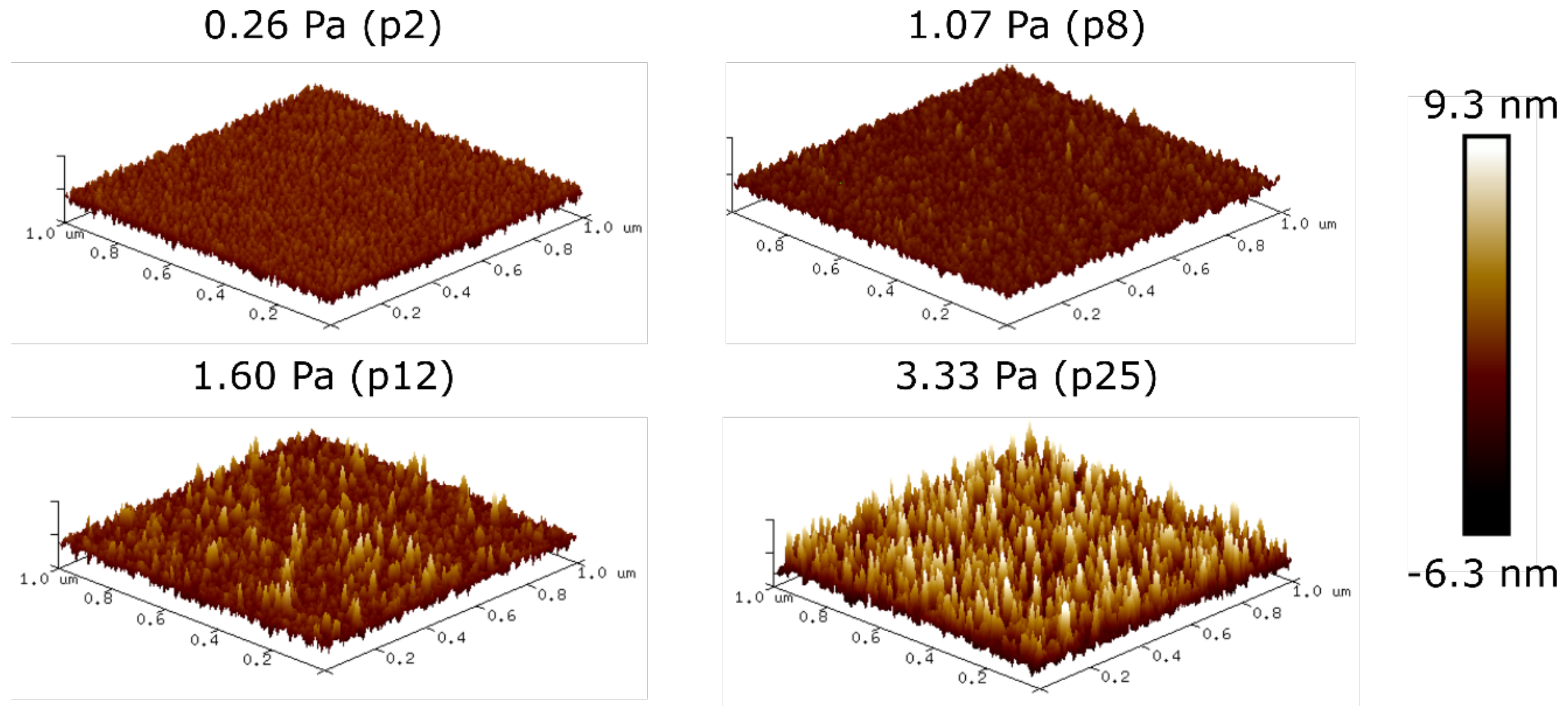

**Figure 2.** AFM height maps (1 $\mu$m × 1 $\mu$m) of Pt thin films sputtered at Ar pressures of 0.27, 1.07, 1.60, and 3.33 Pa, corresponding to samples p2, p8, p12, and p25, respectively. The images reveal a progressive increase in surface roughness with increasing sputter pressure. The color scale is normalized across all images to allow direct comparison of the surface morphology.

XRD measurements provide insight into the structural evolution of the Pt thin films with increasing sputter pressure. All films exhibit a dominant Pt(111) reflection corresponding to the (111) plane of face-centered cubic Pt (fcc, space group Fm-3m) [39]. The measured 2θ position gradually shifts from 39.76° (p1) to 40.09° (p8) and slightly decreases again at higher pressures (Fig. 3a). A weaker Pt(222) reflection is also visible for all samples, with 2θ values ranging from 85.72° to 86.40°. For thin films deposited at sputter pressures higher than 0.53 Pa, an additional weak Pt(020) reflection appears at 2θ ≈ 46.35°. The β-Ta(002) diffraction peak at 2θ ≈ 33.4° (space group: P-421m) [58] confirms the presence of the Ta adhesion layer, while a faint diffraction from the single-crystalline Si substrate is visible in the 2D images (2θ ≈ 70.5°). The β-Ta and Si signals are not resolved in the corresponding 1D diffraction patterns due to their low intensity (Figure S2). Representative 2D diffraction images for p1, p8, and p25 reveal qualitative differences in grain orientation and texture. At low sputter pressure (p1), the Pt(111) reflection appears as a discrete segment of a diffraction arc, indicating strongly textured growth. At intermediate pressure (p8), the Pt(111) arc becomes slightly broader, consistent with an increased angular distribution of grain orientations while maintaining a dominant texture component. For p25, a visible diffraction ring appears together with a brighter Pt(111) intensity segment within the ring, reflecting a mixed growth regime where partially polycrystalline grains coexist with textured grains.

The evolution of the Pt(111) peak position with sputter pressure reflects changes in the out-of-plane lattice strain induced by biaxial residual stress. In-plane tensile stress leads to an out-of-plane lattice contraction, leading to a shift of the diffraction peak toward higher 2θ values, whereas compressive stress produces the opposite effect. In the present thin films, the Pt(111) reflection progressively shifts to higher 2θ values with increasing sputter pressure up to 1.07 Pa (p8), indicating a gradual out-of-plane lattice contraction relative to stress-free bulk Pt [39] (Figure 3b). The first two samples exhibit compressive in-plane strain according to the DHM curvature measurements, which corresponds to an out-of-plane lattice expansion due to the Poisson effect and hence peak positions at lower 2θ values. At higher sputter pressures, the peak position shifts slightly back toward lower 2θ values (p25), indicating a partial stress relaxation, in good agreement with the stress evolution obtained from curvature measurements.

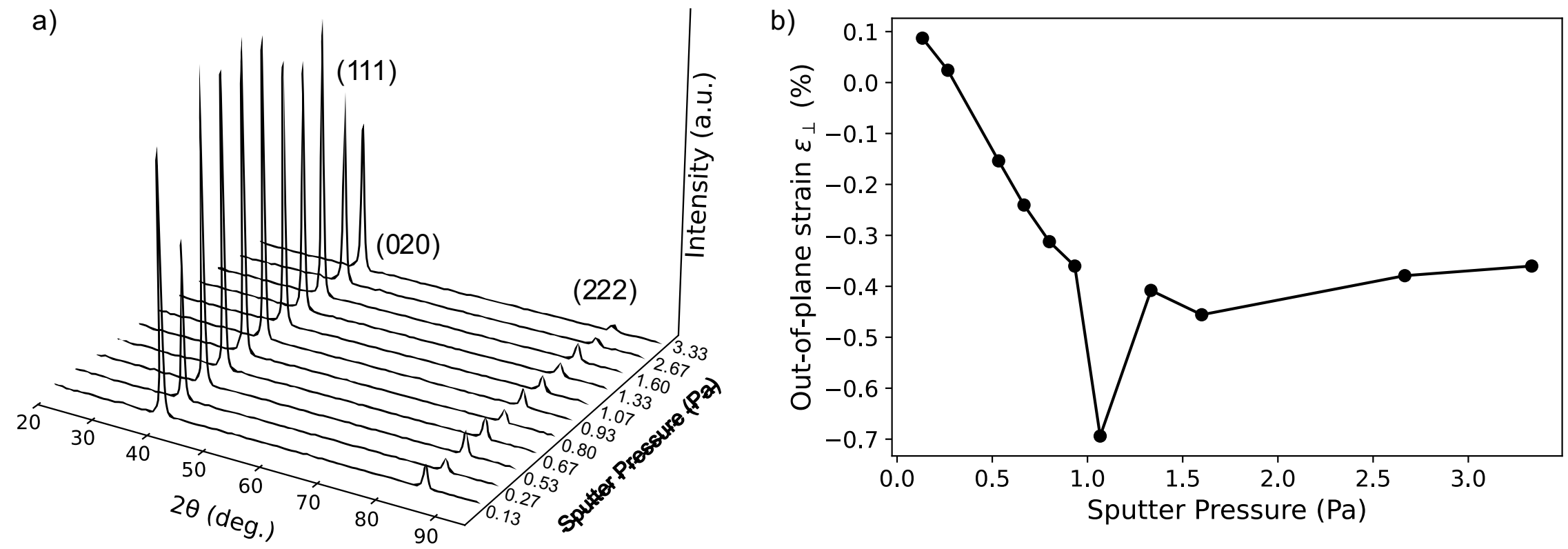

**Figure 3.** (a) XRD patterns of Pt thin films deposited at different sputter pressures shown as a stacked intensity plot to illustrate the evolution of diffraction features with sputter pressure. (b) Average out-of-plane elastic strain $\varepsilon_{\perp}$ of the Pt thin films in dependence of sputter pressure derived from the Pt(111) peak position.

Since electrocatalytic reactions occur at the surface, nanoscale structural variations may influence electrocatalytic activity. However, techniques such as XRD and DHM provide only spatially averaged information on lattice parameters and residual stress, respectively. To further examine lattice parameters at the local scale, cross-sectional TEM combined with SAED analysis was performed. Figure 4 shows a pressure-dependent transition in the Pt thin-film morphology. While films deposited at lower pressure exhibit a comparatively compact columnar structure, increasing the sputter pressure leads to a more open morphology with pronounced intercolumnar gaps. These gaps rarely extend to the substrate, suggesting that a dense basic layer forms during the initial stage of growth before column separation and stress relaxation become dominant. The thickness of this dense layer appears to decrease with increasing sputter pressure. The sample p25 additionally differs in total thickness, being 40% thinner (about 40 nm compared to about 60 nm). The thickness of the Pt thin film and its grain size resulted in a relatively small number of diffraction spots on corresponding SAED patterns. To ensure best possible analysis, each diffraction pattern was rotationally averaged and the interplanar distances were calculated from the corresponding radial intensity profiles using the Digital Micrograph DIFFPack software tool [59]. From each measured interplanar distance, the lattice parameter was calculated, and averaged values are presented in the Supporting Information (Table S1) and Figure 5.

The lattice parameters obtained from cross-sectional samples are consistently larger than the bulk Pt reference value of 0.39237(3) nm [39] for all investigated samples, indicating an out-of-plane lattice expansion. The average lattice parameter decreases from 0.395 nm for the Pt films deposited at low sputter pressures (0.13–0.27 Pa), to 0.394 nm for intermediate pressures (0.93–1.60 Pa), and further to 0.393 nm at higher pressure (3.33 Pa). Considering an experimental uncertainty of ±0.002 nm, this trend indicates a gradual reduction of the out-of-plane lattice expansion with increasing sputter pressure. In contrast to the macroscopic residual stress measured by DHM, which exhibits a non-monotonic dependence on sputter pressure, the lattice strain derived from SAED evolves approximately monotonically within experimental uncertainty. While DHM captures the global mechanical response of the film-substrate system, including stress buildup and partial relaxation, SAED selectively probes the local lattice parameter within nanoscale regions of individual grains. Thus, SAED provides complementary insight into nanoscale lattice strain accommodation at the local crystallographic level, despite the decreasing magnitude of the measured out-of-plane strain at higher deposition pressures. While partial stress relaxation during lamella FIB preparation cannot be excluded, the measured lattice parameters remain well suited for assessing relative trends of local strains between the samples.

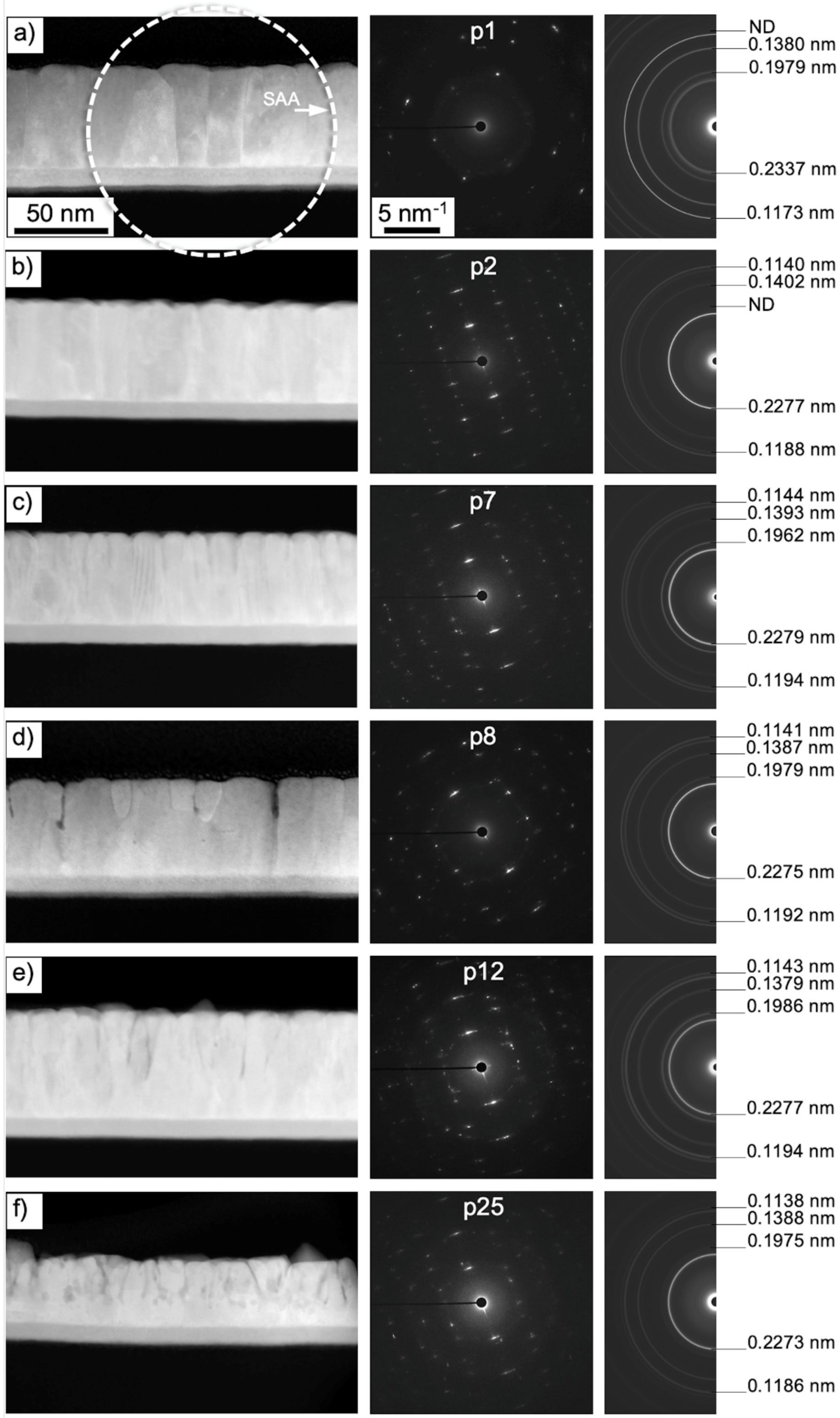

**Figure 4.** TEM assessment of the Pt thin films microstructure and analysis of selected area diffraction patterns (ND – not detected). Patterns were acquired using the smallest available

selected area diffraction aperture (SAA), highlighted in figure a). The right column of the figure shows rotationally averaged patterns with measured interplanar distances for the following planes of the Pt crystal: (111), (200), (220), (311), and (222).

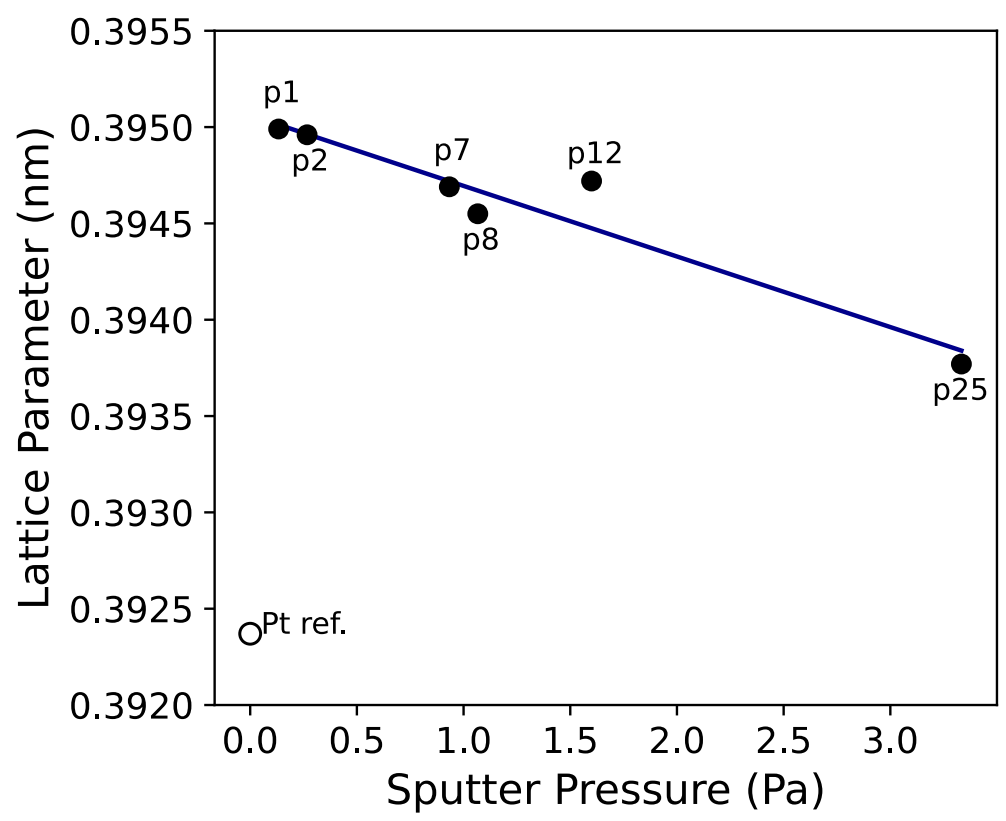

**Figure 5.** Correlation of average lattice parameters and sputter pressure, with a linear fit.

**Correlation between strain and HER Activity**

Representative CVs recorded at a scan rate of 150 $mVs^{-1}$ are shown in Figure 6a. All CVs exhibit characteristic voltammetric features associated with hydrogen adsorption and desorption on Pt surfaces in the potential range of 0.05 - 0.35 V vs. RHE. With increasing residual stress, two desorption peaks at approximately 0.13 and 0.24 V vs. RHE become more pronounced. These features, labeled as D1 and D2, have previously been associated with surface roughening effects on Pt electrodes [60-62]. Specifically, D1 is attributed to Pt(100)-type step sites, while D2 is associated with Pt(111)-type step sites. Surfaces dominated by extended Pt(111) terraces typically show a diminished or absent desorption feature. This observation is consistent with the XRD results, which indicate a stronger (111) texture for films deposited at lower sputter pressure. The ECSA, estimated from the hydrogen adsorption region, is shown in Figure 6b. For films deposited at sputter pressures between 0.13 and 0.93 Pa, the ECSA remains close to the geometric area of the SDC tip (0.002 cm²). At higher pressures (1.07 - 3.3 Pa), the ECSA increases by approximately a factor of three to four. This trend is consistent with AFM analysis, which reveals a threefold increase in surface roughness for pressures above ~ 1.5 Pa.

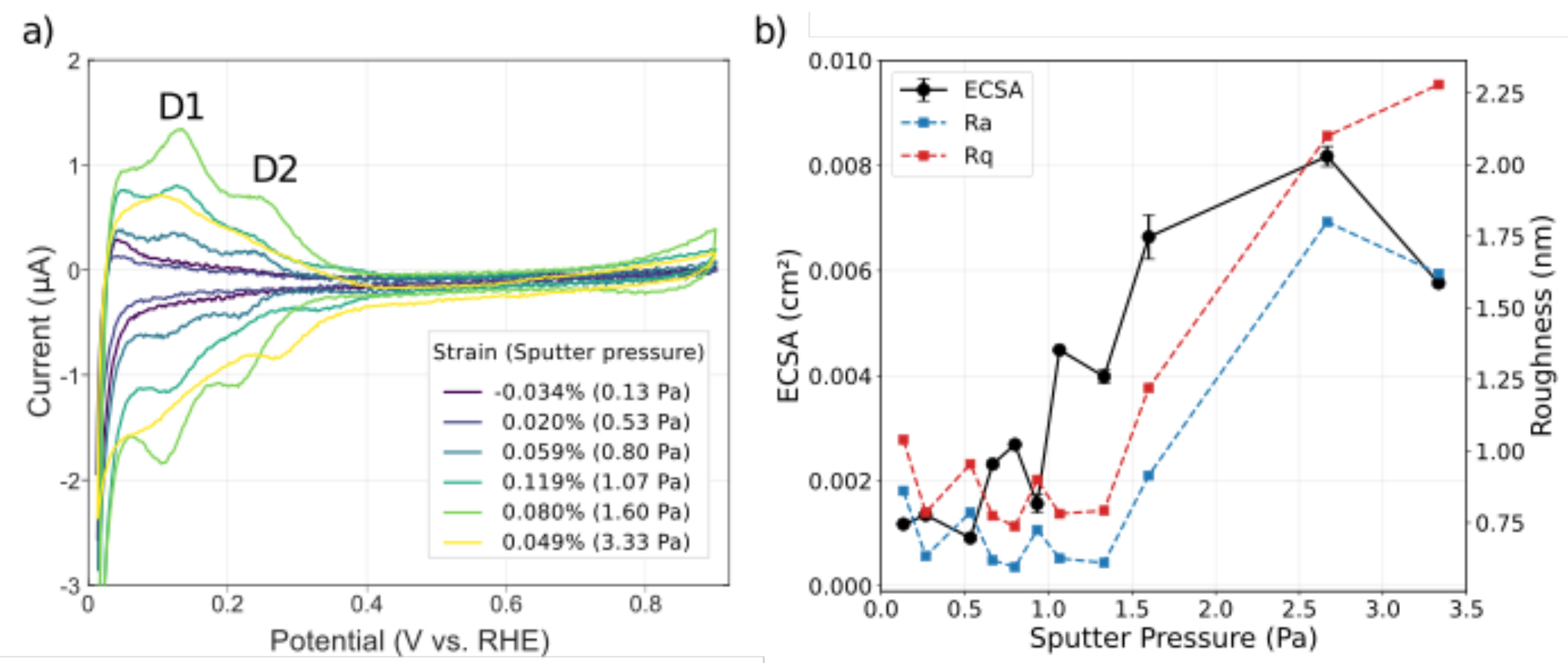

**Figure 6** (a) Cyclic voltammograms of selected Pt thin films illustrating the dependence on hydrogen adsorption and desorption features with sputter pressure. (b) ECSA and AFM roughness parameters in dependence of sputter pressure.

To enable comparison of the HER electrocatalytic activity of the different Pt thin films, the recorded currents during LSVs were normalized to the ECSA. Figure 7 presents potential vs. RHE at a current density of -10 mA/cm² in dependence of residual strain. The data reveal that the HER activity increases as the residual strain shifts from tensile towards compressive values. The thin film deposited at the lowest pressure (0.13 Pa, p1) exhibits the highest HER activity. This film is characterized by compressive strain, pronounced (111) texture, and relatively smooth surface morphology. It has a residual strain of -0.034% and requires a potential vs. RHE of -11.3 mV to reach -10 mA/cm². With increasing sputter pressure, the HER activity decreases, reaching -26.1 mV at -10 mA/cm² for 1.07 Pa (p8), while the residual strain becomes increasingly compressive up to 0.119%. Thin films deposited at even higher pressures however show substantially lower HER activity, down to -48.8 mV vs. RHE at -10 $mA/cm^2$ for 3.33 Pa (p25), despite a reduction in tensile strain (0.049%) and a significant increase in ECSA. This apparent discrepancy indicates that HER activity cannot be rationalized by residual strain alone. While the ECSA remains approximately constant for samples sputtered at lower pressures, it increases substantially at higher pressures, primarily attributed to enhanced surface roughness, resulting in a corresponding decrease in ECSA-normalized activity. These observations demonstrate that sputter pressure simultaneously modifies residual strain and microstructure, particularly surface morphology, leading to a coupled effect that results in the measured HER activity of sputtered thin films. In particular, microstructural factors such as the exposure of less active surface facets and reduced intercolumnar connectivity are likely to contribute to the observed decrease in electrocatalytic activity. Overall, these results indicate that the governing factors for HER activity can be controlled by the sputter pressure. At lower pressures, where ECSA and surface roughness remain approximately constant, variations in activity can be primarily attributed to residual strain. In contrast, at higher pressures, microstructural changes and increased surface roughness dominate the catalytic response. This behavior reflects the multiscale nature of strain in the thin film. While XRD and DHM measurements reveal a non-monotonic, volcano-type change of the macroscopic residual strain - with a tensile maximum at intermediate sputter pressure followed by partial relaxation at higher pressures - SAED shows a continuous decrease in lattice spacing, corresponding to increasing tensile strain at the nanoscale across the entire sputter pressure range. The monotonic nanoscale strain trend revealed by SAED is more consistent with the overall decrease in HER activity than the macroscopic residual strain change, highlighting that the catalytic response is governed by local, surface-sensitive properties. At higher sputter pressures, this nanoscale strain evolution is accompanied with pronounced microstructural changes, indicating that the catalytic response in this regime results from the combined influence of nanoscale strain and microstructural changes.

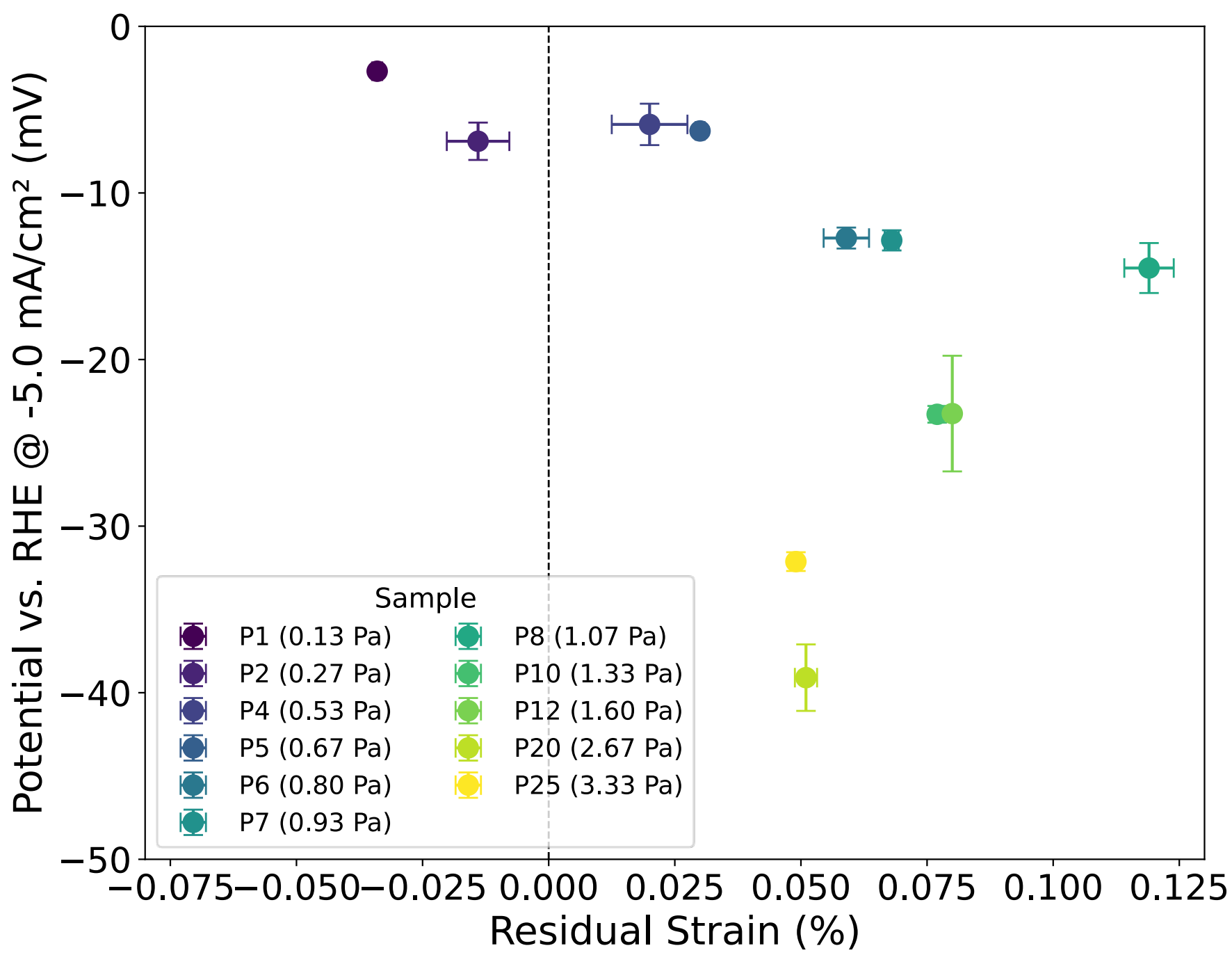

**Figure 7.** HER potential vs. RHE at −10 mA/cm² plotted against residual strain. Currents were normalized to the corresponding ECSA. Error bars represent the standard deviation from repeated LSV measurements at different positions of each sample. Positive strain values correspond to tensile strain, while negative values indicate compressive strain.

## Computational analysis of strain-dependent hydrogen adsorption

Thermodynamics and electrochemical studies indicate that $H_{upd}$ on Pt(111) occupies the 3-fold hollow sites [63,64]. While our calculations yield slightly more negative adsorption energies at top sites, the difference relative to fcc hollow sites is small, on the order of 0.06 eV. Based on experimental observations and to ensure consistency across surfaces, hydrogen adsorption is therefore described at threefold hollow fcc sites throughout this work, in line with previous studies on Pt-like metals [65,66]. For unstrained Pt(111), $\Delta G_H$ of -0.032 eV was obtained, in good agreement with reported literature values [67]. Strain is known to modify adsorption energies by altering metal-metal and metal-adsorbate interactions. By systematically tuning the in-plane lattice constants, tensile and compressive strain provide a controlled means to investigate how deviations from the equilibrium lattice spacing affect hydrogen adsorption and, consequently, HER activity. As demonstrated by Peterson et al. [68], the effect of strain on adsorption energies can be understood in terms of the mechanical response of surface sites to lattice deformation. Within this framework, tensile and compressive strain modify the local atomic environment of the adsorption sites, thereby systematically shifting the adsorption energies. Under tensile strain, adsorption becomes stronger, with adsorption energies of -0.056 eV at 0.5% strain and -0.059 eV at 1% strain. In contrast, compressive strain weakens adsorption, yielding -0.025 eV at 0.5% strain and -0.007 eV at 1% strain. Overall, across the investigated strain range from 1% tensile to 1% compressive strain, hydrogen adsorption energies span from −0.059 eV to −0.007 eV. Beyond these adsorption trends, HER polarization curves further illustrate the electrocatalytic response of thin films. The experimental polarization curves shown in Figure 8a form the basis for the strain-dependent analysis, from which the characteristic potentials are extracted. The curves are ordered by sputter pressure, with the corresponding residual strain values indicated, reflecting the coupled influence of strain and microstructure on the catalytic activity. To isolate the effect of strain, complementary calculated polarization curves (using eq. 4) are presented in Figure 8b.

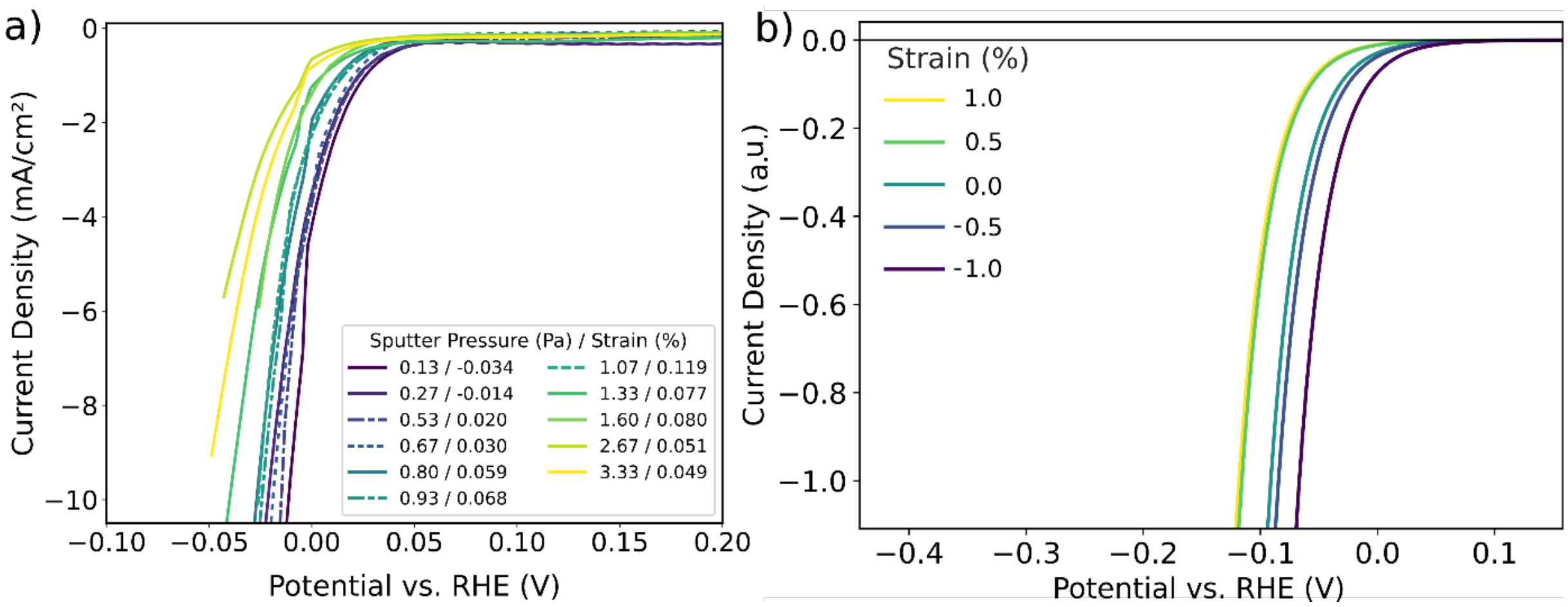

**Figure 8.** (a) Experimental HER polarization curves (LSVs) for Pt thin films deposited at different sputter pressures, corresponding to different residual strain states, showing the current density as a function of potential vs. RHE. Current densities were normalized to the ECSA. (b) Theoretical HER polarization curves for Pt(111) surface under tensile and compressive strain. Compressive strain is indicated according to the sign convention used in the experimental data (negative: compressive; positive: tensile).

The experimental polarization curves reveal a systematic decrease in HER activity with increasing sputter pressure, highlighting the transition from a strain-dominated regime at lower pressures to a microstructure-dominated regime at higher pressures. In contrast, the theoretical results isolate the effect of strain on hydrogen adsorption, which can be rationalized in terms of hydrogen adsorption strength. Compressive strain shifts the adsorption energy closer to the optimal value ($\Delta G_H \approx 0$), enhancing HER activity, whereas tensile strain strengthens hydrogen binding and moves the system away from the Sabatier optimum. The corresponding theoretical polarization curves in Figure 8b capture the same qualitative behavior, showing strain-dependent shifts in HER activity. Consistent with the DFT results, compressive strain weakens hydrogen binding, while tensile strain drives the system into the strong-binding regime. The qualitative agreement between experimental and theoretical trends indicates that strain-dependent changes in adsorption energies captured by DFT contribute to the observed electrocatalytic response. Notably, at higher tensile strain, the DFT results indicate a tendency towards a plateau in HER activity, suggesting that further increase in tensile strain does not significantly alter the adsorption energies. A similar tendency cannot be unambiguously identified in the experimental data, as the observed decrease in activity at higher tensile strain is likely dominated by concurrent microstructural changes, making it challenging to distinguish a potential plateau in the strain-dependent contribution. To further elucidate this behavior, adsorption energies were calculated as a function of hydrogen coverage by consecutively adding hydrogen atoms to the lowest-energy structural configuration up to full coverage. The differential adsorption free energy for the addition of the $n$-th hydrogen atom was calculated as $\Delta G_n = G_n - G_{n-1} - {}^1\!/_2\, G(H_2)$, where $G_n$ represents the free energy of the surface with $n$ adsorbed hydrogen atoms. The integrated free energy is obtained as $G^0 = \Sigma_n \Delta G_n$, representing the cumulative free adsorption energy from zero up to full hydrogen coverage. This procedure yields a discrete set of integral adsorption free energies as a function of hydrogen coverage. To obtain a continuous representation of the free-energy landscape, these data were fitted using a quadratic function, such that:

$$G^0(\theta) = \alpha\theta^2 + \beta\theta \tag{5}$$

Here, the surface coverage is defined as $\theta = n/N_{sites}$, where $N_{sites} = 9$. The parameter $\beta$ corresponds to the adsorption free energy at zero coverage, as defined in equation 3, and $\alpha$ quantifies adsorbate–adsorbate interactions. The linear term describes the intrinsic adsorption energy, while the quadratic term captures coverage-dependent interactions between adsorbed hydrogen atoms. The resulting free-energy profiles for Pt(111) under tensile and compressive strain are shown in Figure 9a.

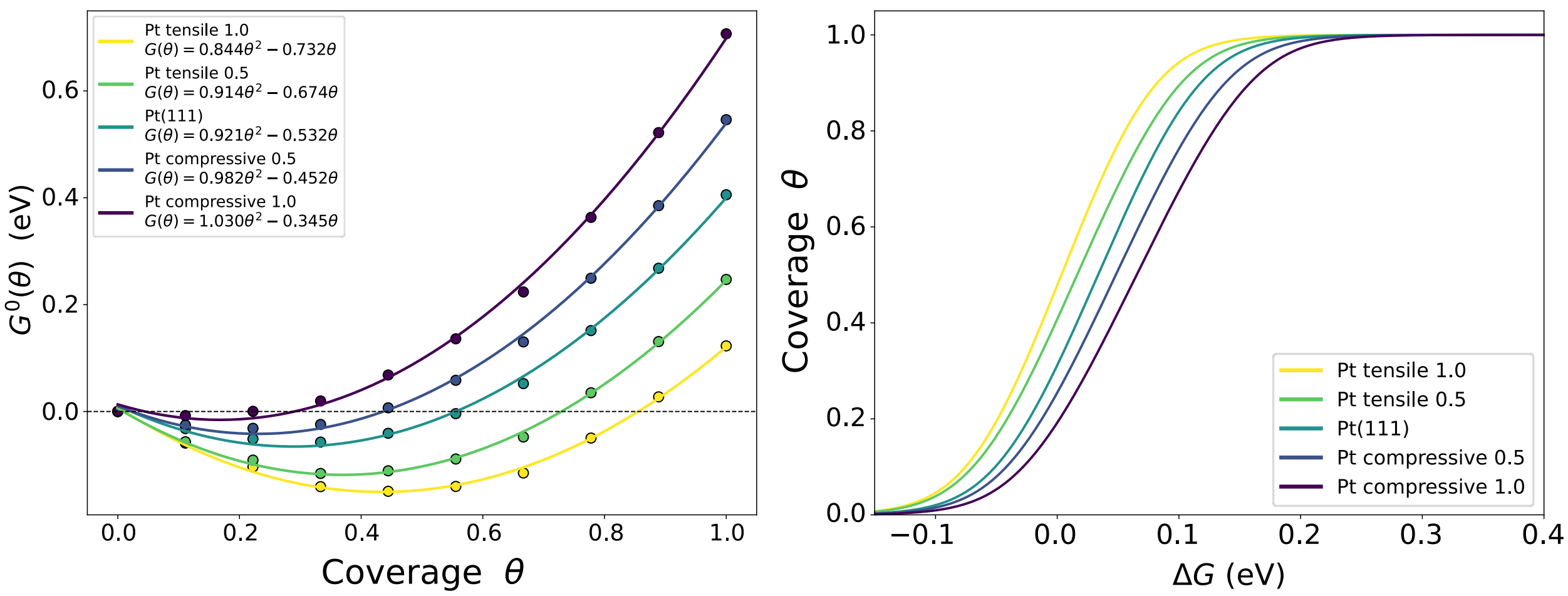

**Figure 9.** a) Quadratic fit of the coverage-dependent hydrogen adsorption free energy, $G^0(\theta)$ (equation 5), on unstrained and strained Pt(111) under tensile and compressive strain. b) Corresponding Frumkin isotherms of hydrogen coverage $\theta$ as a function of total free energy $\Delta G$ (equation 6).

As shown in Figure 9a, the free energy profile depends on the applied strain. The quadratic coefficient $\alpha$ increases progressively from 0.921 eV for Pt(111) to 0.982 eV and 1.030 eV under 0.5% and 1% compressive strain, respectively, indicating enhanced repulsive lateral H–H interactions on the contracted surface. The linear coefficient $\beta$ also shifts systematically from -0.532 eV to -0.452 eV and -0.345 eV, reflecting a progressive weakening of hydrogen adsorption. The opposite trend is observed under tensile strain, where $\beta$ becomes more negative, shifting to -0.674 eV and -0.732 eV, and $\alpha$ reduces to 0.914 eV and 0.844 eV at 0.5% and 1%, respectively, consistent with stronger hydrogen binding and more negative adsorption free energies. The strain-dependent $G^0(\theta)$ profiles modulate hydrogen surface coverage, with tensile strain promoting higher hydrogen surface coverages, consistent with stronger adsorption, while compressive strain reduces coverage and facilitates hydrogen desorption. Although strain induces small variations in $\Delta G_H$, which confine all surfaces close to zero, compressive strain shifts Pt(111) toward $\Delta G_H = 0$. This behavior is reflected in the voltammetric response, as the free energy landscape $G^0(\theta)$ governs the evolution of hydrogen coverage with applied potential. Strain-induced perturbations in $G^0(\theta)$ therefore are captured in the D1 and D2 desorption features. Tensile strain stabilizes adsorbed hydrogen, leading to increased surface coverage and enhanced peak intensities, whereas compressive strain suppresses hydrogen accumulation and diminishes the corresponding voltammetric features. To quantify how strain influences the coverage behavior, the differential adsorption free energy was obtained by differentiating the fitted expression in eq. 5 with respect to the hydrogen coverage $\theta$. Including configurational entropy contributions results to a Frumkin-type adsorption isotherm description the equilibrium hydrogen coverage, where the total free energy can be written as

$$\Delta G = \frac{dG^0}{d\theta} + k_B T \ln\left(\frac{\theta}{1-\theta}\right) \tag{6}$$

The corresponding coverage–free-energy relations obtained from eq. 6 are shown in Figure 9b. The adsorption isotherms exhibit a clear strain dependence. Tensile strain shifts the curves toward more negative $\Delta G$ values, indicating stronger hydrogen adsorption, whereas compressive strain shifts them toward higher $\Delta G$ values, consistent with weaker adsorption. In addition, tensile strain leads to a gradual increase in hydrogen coverage over a wider range of free energies, while compressive strain results in a steeper transition from low to high coverage. This behavior is consistent with the free-energy profiles shown in Figure 9a, where tensile strain results in a broader minimum and compressive strain in a steeper profile.

## CONCLUSIONS

We demonstrate that sputter pressure can be effectively used to tune the residual strain and microstructure of Pt thin films, thereby influencing their HER activity. Residual strain evolves non-monotonically with pressure, consistent with intrinsic stress development during sputter deposition, while increasing pressure simultaneously induces rougher and columnar morphologies. Strain is present across multiple length scales, comprising both film-averaged residual strain and local lattice distortions. While macroscopic measurements reveal a non-monotonic strain evolution with partial saturation, SAED shows a continuous change in lattice spacing at the nanoscale with increasing pressure. HER activity does not follow the macroscopic strain trend but instead decreases at higher sputter pressures. At low pressures, where microstructural variations are limited, activity variations correlate with strain. With increasing pressure, however, morphological changes together with continuously evolving nanoscale lattice distortions coincide with a decline in activity, indicating a transition from a strain-correlated regime to one governed by the combined influence of local strain and microstructural changes. DFT calculations provide a mechanistic framework linking strain to hydrogen adsorption energetics, explicitly accounting for coverage and adsorbate–adsorbate interactions. Strain systematically shifts adsorption energies and alters the coverage-dependent free-energy landscape. Compressive strain progressively weakens hydrogen binding, shifting $\Delta G_H$ toward zero, while tensile strain strengthens adsorption and drives $\Delta G_H$ to more negative values. Despite these variations remaining within a comparable free energy range, the strain-induced changes in $G^0(\theta)$ govern the hydrogen surface coverage, with compressive strain facilitating desorption and tensile strain promoting hydrogen accumulation, trends that suggest compressive strain may provide a more favorable condition for catalytic hydrogen turnover.

Overall, HER activity in sputter-deposited Pt thin films arises from the coupled influence of strain across multiple length scales and pressure-induced microstructural evolution. The observed pressure dependence reflects the interplay of these effects, with strain contributions most evident when structural variations are limited and progressively obscured at higher sputter pressures. These findings establish strain as a key parameter in thin-film electrocatalysis and highlight that its impact can only be understood in the context of the evolving microstructure.

ASSOCIATED CONTENT

**Supporting Information**. The Supporting Information is available free of charge. Additional data are available 10.5281/zenodo.19705685

AUTHOR INFORMATION

**Corresponding Author**

***Alfred Ludwig –** Ruhr-University Bochum, Universitätsstr. 150, 44780 Bochum, Germany; https://orcid.org/0000-0003-2802-6774; Email: alfred.ludwig@rub.de. ***Corina Andronescu –** University Duisburg-Essen, Universitätsstraße 7, 45141 Essen; https://orcid.org/0000-0002-1227-1209; E-Mail: corina.andronescu@uni-due.de.

**Present Addresses**

†Electrochemistry for Energy Storage, Max-Planck-Institut für Chemische Energiekonversion.

**Author Contributions**

The manuscript was written through contributions of all authors. All authors have given approval to the final version of the manuscript.

**Sabrina Baha**: Conceptualization, Methodology, Formal analysis, Validation, Investigation, Data Curation, Writing - Original Draft, Writing - Review & Editing, Visualization; **Alejandro E. Perez Mendoza**: Conceptualization, Methodology, Investigation, Visualization, Validation, Writing - Original Draft, Writing - Review & Editing; **Leonardo Morais**: Formal analysis, Visualization, Investigation, Validation, Writing - Original Draft, Writing - Review & Editing; **Aleksander Kostka**: Investigation, Conceptualization, Visualization, Writing - Original Draft, Writing - Review & Editing ; **Shivam Shukla**: Investigation; **Ellen Suhr**: Investigation, Writing - Review & Editing ; **Andre Oliveira**: Conceptualization, Investigation, Writing - Review & Editing; **Anika Gatzki**: Conceptualization, Investigation; **Henrik H. Kristoffersen**: Writing - Review & Editing; **Jan Rossmeisl**: Methodology; **Corina Andronescu**: Conceptualization, Methodology, Validation, Writing - Review & Editing; **Alfred Ludwig**: Conceptualization, Methodology, Validation, Writing - Review & Editing.

ACKNOWLEDGMENT

The authors acknowledge financial support by the German Research Foundation (Deutsche Forschungsgemeinschaft, DFG) in the framework of the DFG project LU1175/31-1, project AN 1570/2-1 (440951282) and Project-ID 506711657 – SFB 1625, subproject A01 and C02. The ZGH at the Ruhr University Bochum is acknowledged for the use of its equipment (TEM, FIB and XRD).

SUPPORTING INFORMATION

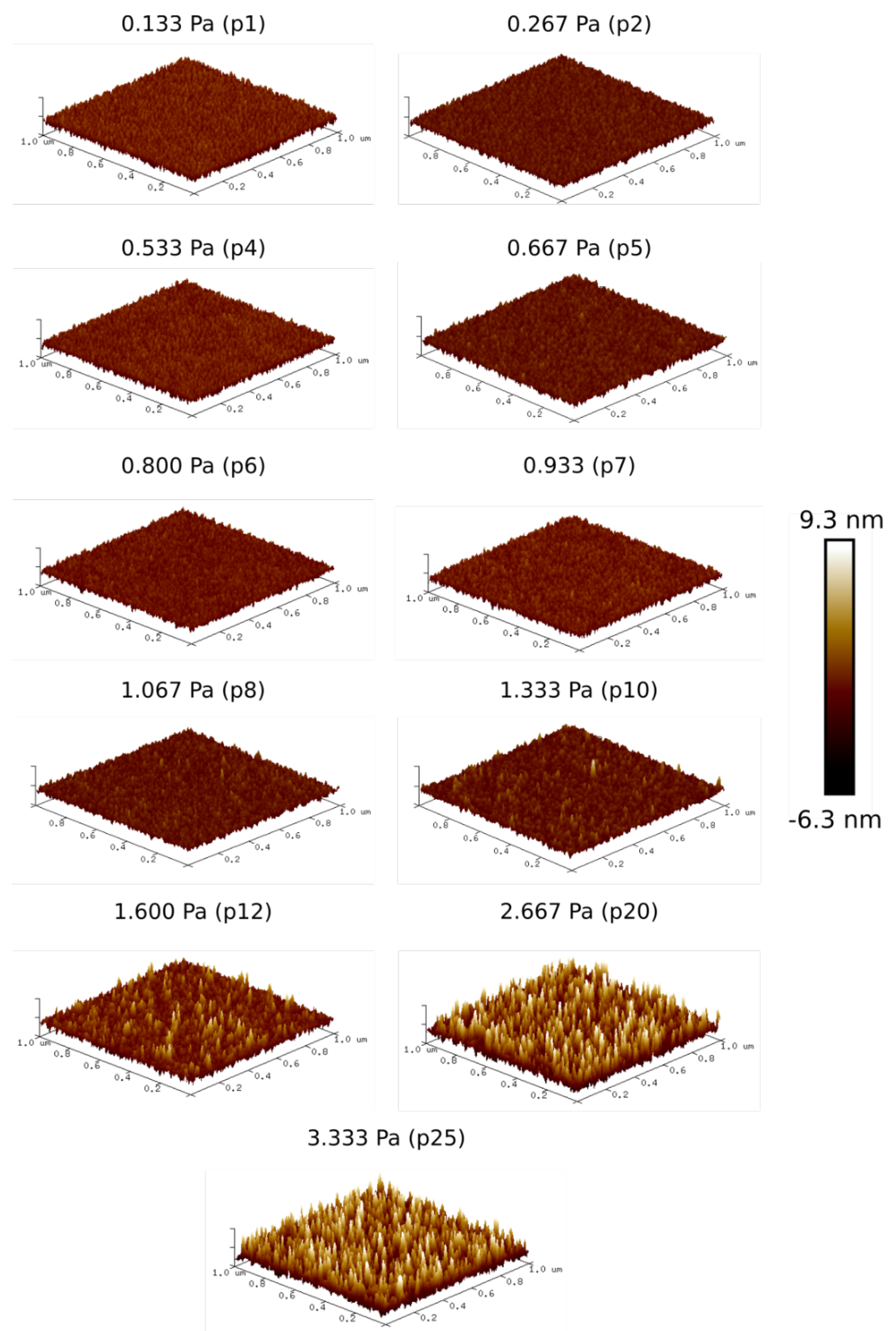

**Figure S1.** AFM 3D images of all Pt thin films deposited with different sputter pressures.

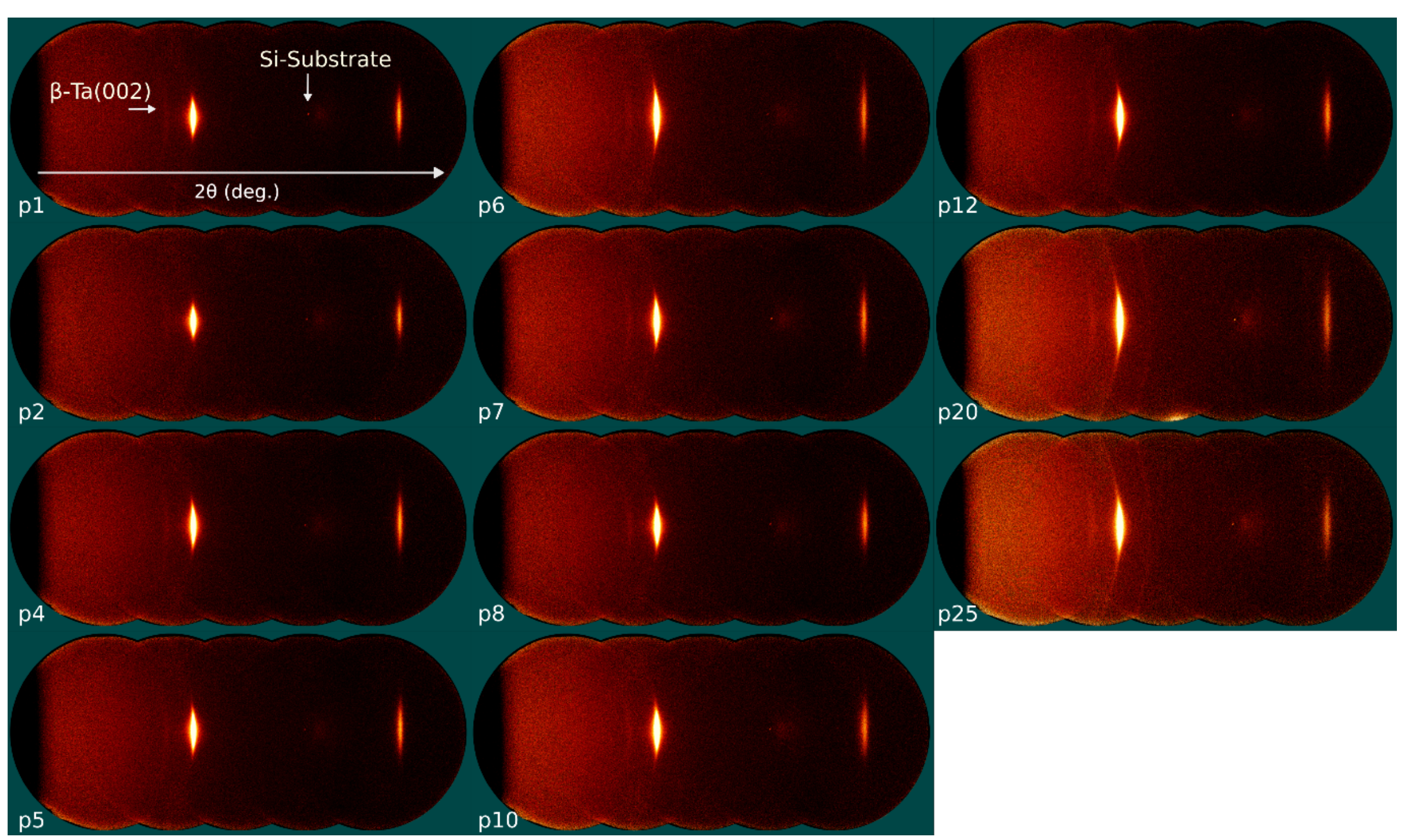

**Figure S2.** Representative 2D diffraction images for samples p1, p8, and p25 showing the transition from strongly textured Pt(111) growth to a mixed regime with both textured and polycrystalline contributions. Weak diffraction features from the β-Ta adhesion layer and the single-crystalline Si substrate are visible in the 2D images.

**Table S1.** Summary of structural and mechanical properties of Pt thin films deposited at different sputter pressures. In-plane stress and strain were obtained by DHM, while out-of-plane lattice spacing and strain were determined by XRD relative to ($d_{111,0}$ = 0.2265) nm (derived from ($a_0$ = 0.39237) nm 54). The lattice parameter $a_{SAED}$ was obtained from SAED.

| Sample | Sputter pressure (Pa) | $\sigma_{\parallel,DHM}$ (MPa) | $\varepsilon_{\parallel,DHM}$ (%) | $d_{111,XRD}$ (nm) | $\varepsilon_{\perp,XRD}$ (%) | $a_{SAED}$ |
|---|---|---|---|---|---|---|
| p1 | 0.133 | -190.1 | -0.16 | 0.2265 | 0.087 | 0.39499 |
| p2 | 0.267 | -76.2 | -0.06 | 0.2264 | 0.024 | 0.39498 |
| p4 | 0.533 | 111.1 | 0.09 | 0.2260 | -0.154 | |
| p5 | 0.667 | 166.9 | 0.14 | 0.2258 | -0.240 | |
| p6 | 0.800 | 325.4 | 0.28 | 0.2256 | -0.312 | |
| p7 | 1.067 | 374.2 | 0.32 | 0.2255 | -0.360 | 0.39428 |
| p8 | 1.066 | 658.6 | 0.56 | 0.2247 | -0.694 | 0.39437 |
| p10 | 1.332 | 426.8 | 0.36 | 0.2254 | -0.408 | |
| p12 | 1.600 | 443.2 | 0.38 | 0.2253 | -0.456 | 0.39441 |
| p20 | 2.667 | 280.0 | 0.24 | 0.2254 | -0.379 | |
| p25 | 3.333 | 295.2 | 0.23 | 0.2255 | -0.360 | 0.39377 |